\documentclass[english,aps,prb,twocolumn]{revtex4-1}
\usepackage[T1]{fontenc}
\usepackage[latin9]{inputenc}
\usepackage{color}
\usepackage{prettyref}
\usepackage{amsmath}
\usepackage{amssymb}
\usepackage{graphicx}
\usepackage{esint}

\makeatletter

\newcommand{\lyxdot}{.}

 \@ifundefined{textcolor}{}
 {%
   \definecolor{BLACK}{gray}{0}
   \definecolor{WHITE}{gray}{1}
   \definecolor{RED}{rgb}{1,0,0}
   \definecolor{GREEN}{rgb}{0,1,0}
   \definecolor{BLUE}{rgb}{0,0,1}
   \definecolor{CYAN}{cmyk}{1,0,0,0}
   \definecolor{MAGENTA}{cmyk}{0,1,0,0}
   \definecolor{YELLOW}{cmyk}{0,0,1,0}
 }

\makeatother

\usepackage{babel}
\begin{document}

\preprint{This line only printed with preprint option}

\title{Generalized Projected Dynamics For Non-System Observables of Nonequilibrium
Quantum Impurity Models}

\author{Guy Cohen}

\affiliation{Department of Chemistry, Columbia University, New York, New York
10027, U.S.A.}

\affiliation{Department of Physics, Columbia University, New York, New York 10027,
U.S.A.}

\author{Eli Y. Wilner}

\affiliation{School of Physics and Astronomy, The Sackler Faculty of Exact Sciences,
Tel Aviv University, Tel Aviv 69978, Israel}

\author{Eran Rabani}

\affiliation{School of Chemistry, The Sackler Faculty of Exact Sciences, Tel Aviv
University, Tel Aviv 69978, Israel}
\begin{abstract}
The reduced dynamics formalism has recently emerged as a powerful
tool to study the dynamics of nonequilibrium quantum impurity models
in strongly correlated regimes. Examples include the nonequilibrium
Anderson impurity model near the Kondo crossover temperature and the
nonequilibrium Holstein model, for which the formalism provides an
accurate description of the reduced density matrix of the system for
a wide range of timescales. In this work, we generalize the formalism
to allow for \textit{non-system} observables such as the current between
the impurity and leads. We show that the equation of motion for the
reduced observable of interest can be closed with the equation of
motion for the reduced density matrix and demonstrate the new formalism
for a generic resonant level model.
\end{abstract}
\maketitle

\section{Introduction\label{sec:Introduction}}

The study of open quantum impurity models, where the coupling of a
small system to multiple baths drives it permanently away from the
possibility of an equilibrium state, is an active and rapidly progressing
field of research. It has recently become possible to make quantitative
statements about experimentally measurable transport properties in
certain cases;\cite{neaton_renormalization_2006,quek_amine-gold_2007,darancet_quantitative_2012}
however, in general many unresolved issues remain. For example, the
nature of the charge and spin dynamics within the Kondo regime of
a quantum dot driven out of equilibrium is currently under investigation,\cite{cohen_numerically_2013}
and basic questions regarding hysteresis and bi-stability in systems
governed by strong electron-phonon couplings remain under debate.\cite{galperin_hysteresis_2004,galperin_non-linear_2008,Bratkovsky2007,Bratkovsky2009,Kosov2011,albrecht_bistability_2012}
Notably, many successful approaches to problems of this kind are based
on master equation treatments and cumulant expansions,\cite{li_quantum_2005,li_spontaneous_2005,luo_calculation_2007,jin_exact_2008,li_improved_2011}
or on diagrammatic partial summations,\cite{konig_resonant_1995}
all of which are approximate in general. A major theoretical challenge
lies in the need to provide an accurate account of time propagation
of open quantum systems, starting from some known initial state and
proceeding all the way to an unknown steady-state.

Numerically exact methods play a particularly important role in the
quest to obtain a reliable, unbiased description of nonequilibrium
phenomena. Several different types of brute-force approaches developed
in recent years have been applied to open nonequilibrium quantum systems.
These include the time-dependent numerical renormalization group\cite{anders_real-time_2005}
and functional renormalization group,\cite{jakobs_nonequilibrium_2007,kennes_renormalization_2012,kennes_oscillatory_2013}
time-dependent density matrix renormalization group,\cite{white_density_1992,schmitteckert_nonequilibrium_2004,dias_da_silva_transport_2008,heidrich-meisner_real-time_2009}
iterative~\cite{weiss_iterative_2008,eckel_comparative_2010,segal_numerically_2010-1,hutzen_iterative_2012}
and stochastic~\cite{muhlbacher_real-time_2008-1,werner_diagrammatic_2009,schiro_real-time_2009,werner_weak-coupling_2010,gull_bold-line_2010}
diagrammatic methods, and wavefunction based approaches.\cite{wang_numerically_2009,wang_numerically_2011}
While the application of these approaches to the the nonequilibrium
Holstein, the Anderson impurity, and the spin-fermion models has been
very fruitful, they are still restricted to a relatively small range
of parameters, typically characterized by a rapid decay to steady-state.
Situations or observables exhibiting slow dynamics are inaccessible
by these brute-force methods.

An alternative approach recently proposed by Cohen and Rabani~\cite{cohen_memory_2011-1}
is based on a combination of a brute-force impurity solver (one of
the above) with a generalized quantum master equation (GQME). The
Nakajima--Zwanzig--Mori~\cite{nakajima_quantum_1958,zwanzig_ensemble_1960,mori_transport_1965}
formalism was used to derive an exact equation of motion for the reduced
density matrix of the system, which includes a memory kernel giving
rise to non-Markovian effects. This kernel, along with some information
regarding the initial conditions, determines the dynamics of the system
and contains all information about the time dependence of single-time
system observables and their steady-state values. In many situations
of interest, in particular when the bandwidth of the baths is large
compared to other energy scales in the problem, the memory kernel
is expected to decay rapidly to zero.\cite{weiss_iterative_2008,cohen_memory_2011-1,cohen_numerically_2013,Wilner2013}
Thus, one can safely truncate the memory kernel at a finite time,
performing a ``cutoff approximation''. Brute-force impurity solvers
limited to short times are well suited for the kernel's numerical
evaluation up to the cutoff time, and once the memory kernel has been
obtained, the GQME is exact and tractable at \textit{all} times.

The GQME formalism has recently been combined with the Bold impurity
solver~\cite{gull_bold-line_2010,gull_numerically_2011} to uncover
the spin dynamics near the Kondo crossover temperature~\cite{cohen_numerically_2013}
and with the multilayer multi-configuration time-dependent Hartree
method to reveal the nature of bi-stability in systems with electron-phonon
couplings.\cite{Wilner2013} Despite the open nature of the systems
studied in these works, transport properties were not addressed; this
is due to one of the formalism's main limitations, in that observables
outside the impurity part of the Hilbert space are not accessible,
and only system observables such as the dot's population or magnetization
can be calculated. On the other hand, perturbative expressions for
transport properties in terms of vertex functions have been derived
and evaluated before in approximate methodologies built on the GQME,\cite{leijnse_kinetic_2008,schoeller_perturbative_2009-1}
and it seems reasonable to expect that a general exact formulation
in the spirit of Ref. \cite{cohen_memory_2011-1} should exist.

In this paper we extend the GQME formalism (reviewed in \prettyref{sec:Projected-dynamics-system})
to describe non-system observables. This allows for comparison of
predictions made by the GQME with a much wider variety of experimental
observables, of which an important example (worked out in detail here)
is the current. Equally important, it facilitates access to the spectral
functions by way of measuring the current to an infinitesimally coupled
auxiliary bath~\cite{sun_kondo_2001,lebanon_measuring_2001,muhlbacher_anderson_2011}.
The key idea discussed in \prettyref{sec:Generalized-projected-dynamics}
is based on deriving a reduced equation of motion for the observable
of interest, which can then be expressed in terms of the reduced density
matrix. This leads to an introduction of an additional, observable-specific
memory kernel with properties qualitatively similar to those of the
memory kernel appearing in the standard GQME. \prettyref{sec:Steady-State}
is devoted to expressing the steady-state properties in terms of the
memory kernels alone, while in \prettyref{sec:Expressing-kernels}
we show how the projected quantities appearing in the GQME can be
translated into the language of ordinary observables expressed as
second-quantized operators, using a noninteracting model as an illustrative
example. In \prettyref{sec:Results} we present several test cases
and examples for the non-interacting case, where the properties of
memory kernels can be explored without the need for technically complicated
numerical solvers. Finally, a summary is given in \prettyref{sec:Summary-and-conclusions}.

\section{Projected dynamics for system observables\label{sec:Projected-dynamics-system}}

We will begin by reviewing the derivation of exact projected (or reduced)
equations of motion,\cite{zwanzig_nonequilibrium_2001} and the process
of going from projected to unprojected dynamics.\cite{zhang_nonequilibrium_2006}
These details are provided here in a self-contained manner because
they will be of particular importance later, when we discuss how the
process can be generalized. Consider an operator Hilbert space $\mathcal{H}=\mathcal{S}\otimes\mathcal{B}$
composed of two subspaces $\mathcal{S}$ and $\mathcal{B}$, which
we will call the \emph{system} and \emph{bath} subspaces. We are interested
in a Hamiltonian of the form
\begin{equation}
H=H_{S}+H_{B}+V,
\end{equation}
where $H_{S}\in\mathcal{S}$ is the system or impurity Hamiltonian,
$H_{B}\in\mathcal{B}$ is the bath Hamiltonian and $V\in\mathcal{H},\, V\notin\mathcal{S},\,\mathcal{B},$
is the coupling Hamiltonian. Generally, the motivation for employing
such a description is to describe a small, strongly interacting impurity
coupled to large noninteracting baths, but we need make no further
assumptions at this stage. We can now define a projection operator
$P$ onto the $\mathcal{S}$ subspace by tracing out the bath degrees
of freedom, in the process also defining its complementary operator
$Q$:
\begin{eqnarray}
P & = & \rho_{B}\mathrm{Tr}_{B},\\
Q & = & 1-P.
\end{eqnarray}
Here $\rho_{B}=e^{-\beta H_{B}}/\mathrm{Tr}_{B}e^{-\beta H_{B}}$.
We also define $\rho_{S}\in\mathcal{S}$ to be the initial impurity
density matrix, and $\rho_{0}=\rho_{B}\otimes\rho_{S}$ the initial
full density matrix. The expectation value of a system operator $A\in\mathcal{S}$
is given by 
\begin{eqnarray}
\left\langle A\left(t\right)\right\rangle  & = & \mathrm{Tr}\rho\left(t\right)A\\
 & = & \mathrm{Tr}_{S}\left[\left(\mathrm{Tr}_{B}\rho\left(t\right)\right)A\right]\\
 & \equiv & \mathrm{Tr}_{S}\left\{ \sigma\left(t\right)A\right\} ,\label{eq:system_operator_EOM}
\end{eqnarray}
and the reduced density matrix $\sigma\left(t\right)=\mathrm{Tr}_{B}\rho\left(t\right)$
contains information about all single-time properties of system observables.
This object has a lower dimensionality than that of $\rho$, and it
would thus be economical to describe its equations of motion without
referring to the system as a whole. This is the basic idea behind
reduced quantum dynamics.

To proceed, one considers the Liouville--von Neumann equation, which
governs the dynamics of the full density matrix:
\begin{eqnarray}
i\hbar\frac{\mathrm{d}}{\mathrm{d}t}\rho & =\left[H,\rho\right]\equiv & \mathcal{L}\rho.
\end{eqnarray}
The Liouvillian superoperator $\mathcal{L}$ denotes performing a
commutation with the Hamiltonian, such that $\mathcal{L}A\equiv\left[H,A\right]$.
We also define $\mathcal{L}_{S}A\equiv\left[H_{S},A\right]$, $\mathcal{L}_{V}A\equiv\left[V,A\right]$
and $\mathcal{L}_{B}A\equiv\left[H_{B},A\right]$. Applying each of
the projection operators from the left and using $1=P+Q$ within the
commutator gives:
\begin{eqnarray}
i\hbar\frac{\mathrm{d}}{\mathrm{d}t}P\rho & = & P\left[H,\left(P+Q\right)\rho\right],\label{eq:density_P_part}\\
i\hbar\frac{\mathrm{d}}{\mathrm{d}t}Q\rho & = & Q\left[H,\left(P+Q\right)\rho\right].\label{eq:density_Q_part}
\end{eqnarray}
Eq.~\prettyref{eq:density_Q_part} has the formal solution
\begin{eqnarray}
Q\rho & = & e^{-\frac{i}{\hbar}Q\mathcal{L}t}Q\rho_{0}\nonumber \\
 &  & \,-\frac{i}{\hbar}\int_{0}^{t}\mathrm{d}\tau\, e^{-\frac{i}{\hbar}Q\mathcal{L}\tau}Q\mathcal{L}\rho_{B}\sigma\left(t-\tau\right),\label{eq:Qrho_formal_solution}
\end{eqnarray}
which can be inserted into Eq.~\prettyref{eq:density_P_part} to
obtain the Nakajima--Zwanzig--Mori equation (NZME)\cite{nakajima_quantum_1958,zwanzig_ensemble_1960,mori_transport_1965}:
\begin{eqnarray}
i\hbar\dot{\sigma}\left(t\right) & = & \mathcal{L}_{S}\sigma\left(t\right)+\vartheta\left(t\right)-\frac{i}{\hbar}\int_{0}^{t}\mathrm{d}\tau\kappa\left(\tau\right)\sigma\left(t-\tau\right),\label{eq:sigma_eom}\\
\kappa\left(t\right) & \equiv & \mathrm{Tr}_{B}\left\{ \mathcal{L}_{V}e^{-\frac{i}{\hbar}Q\mathcal{L}t}Q\mathcal{L}\rho_{B}\right\} ,\label{eq:kappa_definition}\\
\vartheta\left(t\right) & \equiv & \mathrm{Tr}_{B}\left\{ \mathcal{L}_{V}e^{-\frac{i}{\hbar}Q\mathcal{L}t}Q\rho_{0}\right\} .\label{eq:theta_definition}
\end{eqnarray}

Let us take a moment to examine the important relation of Eq.~\prettyref{eq:sigma_eom}.
It has the form of an operator linear Volterra integro-differential
equation of the second kind. As we have made no approximations, it
is exact; yet it contains only operators and superoperators within
the low-dimensional system space. The time derivative of the reduced
density matrix $\sigma$ is given by the sum of three contributions:
the first term ($\mathcal{L}_{S}\sigma\left(t\right)$) describes
the exact evolution of the system if the coupling to the bath were
set to zero. The second term ($\vartheta\left(t\right)$) expresses
initial correlations between the system and bath, and it is easy to
verify from its definition in Eq.~\prettyref{eq:theta_definition}
that it equals zero for the factorized initial conditions $\rho_{0}=\rho_{B}\otimes\rho_{S}$
(we will assume this later, but keep this term for generality, as
access to general initial conditions is of some interest when considering,
for instance, quenching). The last term includes the memory kernel
($\kappa\left(\tau\right)$), and depends on the complete history
of $\sigma\left(t\right)$ at earlier times. The appearance of this
non-Markovian term is the price of going to reduced dynamics, and
to make headway with the NZME one must begin by evaluating $\kappa\left(t\right)$.

The definition of $\kappa\left(t\right)$ in Eq.~\prettyref{eq:kappa_definition}
includes the troublesome component $e^{-\frac{i}{\hbar}Q\mathcal{L}\tau}$.
To understand why it is troubling, consider the following: it is easy
to show that the superoperator $e^{-\frac{i}{\hbar}\mathcal{L}\tau}$
evolves the density matrix with respect to the Hamiltonian, thus simply
expressing our familiar notion of dynamics:
\begin{equation}
e^{-\frac{i}{\hbar}i\mathcal{L}\tau}\rho=e^{\frac{i}{\hbar}H\tau}\rho e^{-\frac{i}{\hbar}H\tau}.
\end{equation}
The modified operator $e^{-\frac{i}{\hbar}Q\mathcal{L}\tau}$, however,
contains a projection operator within the exponent, and so does something
else entirely---something which turns out to be substantially harder
to understand or calculate. Our next step is therefore to get rid
of these inconvenient projected dynamics. While several ways to go
about this task exist, we will limit the discussion to a particular
method suggested by Zhang \textit{et al}.\cite{zhang_nonequilibrium_2006}

Consider the function $\vartheta\left(t\right)$ of Eq.~\prettyref{eq:theta_definition}.
By applying the identity
\begin{equation}
e^{-\frac{i}{\hbar}Q\mathcal{L}t}=e^{-\frac{i}{\hbar}\mathcal{L}t}+\frac{i}{\hbar}\int_{0}^{t}\mathrm{d}\tau e^{-\frac{i}{\hbar}\mathcal{L}\left(t-\tau\right)}P\mathcal{L}e^{-\frac{i}{\hbar}Q\mathcal{L}\tau}\label{eq:projection_identity}
\end{equation}
to its definition, we can obtain:
\begin{eqnarray}
\vartheta\left(t\right) & = & \mathrm{Tr}_{B}\left\{ \mathcal{L}_{V}e^{-\frac{i}{\hbar}Q\mathcal{L}t}Q\rho_{0}\right\} \\
 & = & \mathrm{Tr}_{B}\Bigg\{\mathcal{L}_{V}e^{-\frac{i}{\hbar}\mathcal{L}t}Q\rho_{0}\nonumber \\
 &  & +\frac{i}{\hbar}\int_{0}^{t}\mathrm{d}\tau\mathcal{L}_{V}e^{-\frac{i}{\hbar}\mathcal{L}\left(t-\tau\right)}P\mathcal{L}e^{-\frac{i}{\hbar}Q\mathcal{L}\tau}Q\rho_{0}\Bigg\}\\
 & = & \Xi\left(t\right)-\Phi\left(t\right)\sigma\left(0\right)\nonumber \\
 &  & +\frac{i}{\hbar}\int_{0}^{t}\mathrm{d}\tau\Phi\left(t-\tau\right)\vartheta\left(\tau\right),\label{eq:theta_EOM}
\end{eqnarray}
where
\begin{eqnarray}
\Xi\left(t\right) & = & \mathrm{Tr}_{B}\left\{ \mathcal{L}_{V}e^{-\frac{i}{\hbar}\mathcal{L}t}\rho_{0}\right\} ,\label{eq:Xi_definition}\\
\Phi\left(t\right) & = & \mathrm{Tr}_{B}\left\{ \mathcal{L}_{V}e^{-\frac{i}{\hbar}\mathcal{L}t}\rho_{B}\right\} .\label{eq:Phi_definition}
\end{eqnarray}
Applying the same identity Eq.~\prettyref{eq:projection_identity}
to Eq.~\prettyref{eq:kappa_definition} yields:
\begin{equation}
\kappa\left(t\right)=i\hbar\dot{\Phi}\left(t\right)-\Phi\left(\tau\right)\mathcal{L}_{S}+\frac{i}{\hbar}\int_{0}^{t}\mathrm{d}\tau\Phi\left(t-\tau\right)\kappa\left(\tau\right).\label{eq:kappa_EOM}
\end{equation}

Eqs.~\prettyref{eq:theta_EOM} and \prettyref{eq:kappa_EOM} are
superoperator linear Volterra integral equations of the second kind,
with both the inhomogeneous contributions and the kernels determined
by the combination of Eq.~\prettyref{eq:Xi_definition} and \prettyref{eq:Phi_definition}
and the form of the system Liouvillian operator. Like the NZME, Eq.~\ref{eq:sigma_eom},
they consist of objects which inhabit the low-dimensional impurity
subspace---however, they have a higher dimensionality due to their
superoperator nature (if $\sigma$ can be represented by an $N\times N$
matrix, then $\vartheta$ and $\kappa$ are $N^{2}\times N^{2}$).
Their importance lies in the fact that $\Phi$ and $\Xi$, which are
propagated by the full Hamiltonian with normal dynamics, can be written
in terms of physical observables; this means they can be evaluated
with a variety of computational methods, and then used to solve Eqs.~\prettyref{eq:theta_EOM}
and \prettyref{eq:kappa_EOM} numerically.

We now have the necessary machinery at hand to introduce the \emph{cutoff
approximation}: if we have some way of evaluating $\kappa\left(t\right)$
up to some finite time, it is sometimes possible to make an ansatz
about later times. Importantly, if the memory has decayed to zero
to within a numerical accuracy over a finite time, one can assume
that it will remain zero at all later times. One then solves the NZME
with this cutoff memory kernel to obtain an approximate value for
$\sigma\left(t\right)$; however, if $\sigma(t)$ can be converged
in the cutoff time to within the desired accuracy, the entire procedure
is numerically exact. Note that while in principle this procedure
can be performed for any Hamiltonian (regardless of the form of the
interactions), for it to be beneficial in practice the system in question
should exhibit dynamical timescales substantially longer than those
of the memory decay time.

\section{Generalized projected dynamics for non-system observables\label{sec:Generalized-projected-dynamics}}

The time-dependent electronic current flowing through an impurity
does not have a single definition, as it depends in general on the
topology of the surface through which electronic flow is measured.
In steady state populations must be constant, and the current must
therefore become independent of this definition (if it is unique);
however, the definition itself remains arbitrary. This is well known
and usually does not warrant much discussion, yet in the context of
reduced dynamics a subtle point occurs: if the impurity model in question
is, for instance, given by a chain Hamiltonian, current may be measured
at any point along the chain and may be obtained from knowledge of
$\sigma\left(t\right)$. However, in models where not all current
must flow between impurity sites, it is necessary to measure currents
at the junction between the impurity and one of the leads(baths).
In a setup involving two Fermionic baths held at different chemical
potentials, often referred to as the left ($L$) and right ($R$)
\emph{leads}, one is therefore interested in quantities such as the
so-called ``left current'' (or alternatively the ``right current''):
\begin{equation}
I_{L}\equiv\frac{\mathrm{d}}{\mathrm{d}t}eN_{L}=\sum_{q\in L}\left\langle \frac{ie}{\hbar}\left[H,a_{q}^{\dagger}a_{q}\right]\right\rangle .
\end{equation}
Here the $a_{k}^{\dagger}$ and $a_{k}$ are creation and destruction
operators in the left lead subspace $L\subset\mathcal{B}$. The current
operator will therefore in general \emph{not} be a member of the impurity
subspace $\mathcal{S}$, and \emph{cannot} be obtained from Eq.~\prettyref{eq:system_operator_EOM}
along with knowledge of the reduced density matrix $\sigma\left(t\right)$.
It should be mentioned briefly that one simple way of dealing with
this issue is to define an effective Hamiltonian and a repartitioning
of $\mathcal{H}$ in such a way that the current can be measured within
the system; however, this will invariably raise the dimensionality
of $\mathcal{S}$, which may complicate the problem beyond the applicability
of many numerical methods.

In order to allow for the calculation of non-system observables, we
proceed by deriving a Nakajima--Zwanzig--Mori-like equation for the
expectation value of a general operator $I$. While $I$ implies that
we are interested in the current, nothing in this section is limited
to that specific case. The ideas to follow could work equally well
for any operator, but for the current one might expect them to converge
quickly with a cutoff time, since it is expected to be determined
largely by quantities local to the dot.

We start from the equation of motion for $I$ in the Schrödinger picture,
where $\rho$ has a time dependence but $I$ does not:
\begin{equation}
i\hbar\frac{\mathrm{d}}{\mathrm{d}t}I\rho=I\left[H,\rho\right].
\end{equation}
Applying the projection operators using the definitions and procedure
of the previous section yields:
\begin{eqnarray}
i\hbar\frac{\mathrm{d}}{\mathrm{d}t}PI\rho & = & PI\mathcal{L}P\rho+PI\mathcal{L}Q\rho,\label{eq:current_P_part}\\
i\hbar\frac{\mathrm{d}}{\mathrm{d}t}QI\rho & = & QI\mathcal{L}P\rho+QI\mathcal{L}Q\rho.\label{eq:current_Q_part}
\end{eqnarray}
In addition to these, we still have equations \prettyref{eq:density_P_part}
and \prettyref{eq:density_Q_part} for the density matrix, which may
be written in the form:
\begin{eqnarray}
i\hbar\frac{\mathrm{d}}{\mathrm{d}t}P\rho & = & P\mathcal{L}P\rho+P\mathcal{L}Q\rho,\label{eq:sigma_P_part}\\
i\hbar\frac{\mathrm{d}}{\mathrm{d}t}Q\rho & = & Q\mathcal{L}P\rho+Q\mathcal{L}Q\rho.\label{eq:sigma_Q_part}
\end{eqnarray}
The latter equation, as before, has the formal solution Eq.~\prettyref{eq:Qrho_formal_solution}.
Putting this expression together with Eq.~\prettyref{eq:current_P_part}
and defining $\iota\left(t\right)=\mathrm{Tr}_{B}I\rho$, we obtain
(with $\dot{\iota}\equiv\frac{\mathrm{d}\iota}{\mathrm{d}t}$)
\begin{eqnarray}
i\hbar\dot{\iota}\left(t\right) & = & \mathrm{Tr}_{B}I\mathcal{L}\rho_{B}\sigma\left(t\right)+\mathrm{Tr}_{B}I\mathcal{L}e^{-\frac{i}{\hbar}Q\mathcal{L}t}Q\rho_{0}\\
 &  & \,-\frac{i}{\hbar}\int_{0}^{t}\mathrm{d}\tau\,\mathrm{Tr}_{B}\left\{ I\mathcal{L}e^{-\frac{i}{\hbar}Q\mathcal{L}\left(t-\tau\right)}Q\mathcal{L}\rho_{B}\right\} \sigma\left(\tau\right),\nonumber 
\end{eqnarray}
or
\begin{equation}
i\hbar\dot{\iota}\left(t\right)=\mathcal{L}_{\iota}\sigma\left(t\right)+\vartheta_{\iota}\left(t\right)-\frac{i}{\hbar}\int_{0}^{t}\mathrm{d}\tau\,\kappa_{\iota}\left(t-\tau\right)\sigma\left(\tau\right).\label{eq:iota_eom}
\end{equation}
The terms of this equation appear similar to those of the NZME in
Eq.~\prettyref{eq:sigma_eom}, though it is a solution in closed
form rather than an integro-differential equation, since $\iota\left(t\right)$
appears only on the left hand side. In writing it we have defined:
\begin{eqnarray}
\mathcal{L}_{\iota} & \equiv & \mathrm{Tr}_{B}\left\{ I\mathcal{L}\rho_{B}\right\} ,\\
\vartheta_{\iota}\left(t\right) & \equiv & \mathrm{Tr}_{B}\left\{ I\mathcal{L}e^{-\frac{i}{\hbar}Q\mathcal{L}t}Q\rho_{0}\right\} ,\\
\kappa_{\iota}\left(t\right) & \equiv & \mathrm{Tr}_{B}\left\{ I\mathcal{L}e^{-\frac{i}{\hbar}Q\mathcal{L}t}Q\mathcal{L}\rho_{B}\right\} .
\end{eqnarray}

The initial correlation term $\vartheta_{\iota}$ is once again zero
for uncorrelated initial conditions and this time we will remove it
for the sake of brevity. As in the formalism for $\sigma$, in order
to phrase everything in terms of quantities with unprojected dynamics
we now once again perform the Zhang--Ka--Geva transformation\cite{zhang_nonequilibrium_2006}
on the current memory kernel $\kappa_{\iota}$. Applying the identity
\prettyref{eq:projection_identity} allows us to write
\begin{eqnarray}
\kappa_{\iota}\left(t\right) & = & \mathrm{Tr}_{B}\left\{ I\mathcal{L}e^{-\frac{i}{\hbar}\mathcal{L}t}Q\mathcal{L}\rho_{B}\right\} \\
 &  & +\frac{i}{\hbar}\int_{0}^{t}\mathrm{d}\tau\mathrm{Tr}_{B}\left\{ I\mathcal{L}e^{-\frac{i}{\hbar}\mathcal{L}\left(t-\tau\right)}P\mathcal{L}e^{-\frac{i}{\hbar}Q\mathcal{L}\tau}Q\mathcal{L}\rho_{B}\right\} \nonumber \\
 & = & i\hbar\dot{\Phi}_{\iota}\left(t\right)-\Phi_{\iota}\left(t\right)\mathcal{L}_{S}\nonumber \\
 &  & +\frac{i}{\hbar}\int_{0}^{t}\mathrm{d}\tau\Phi_{\iota}\left(t-\tau\right)\kappa\left(\tau\right),
\end{eqnarray}
or
\begin{eqnarray}
\kappa_{\iota}\left(t\right) & = & i\hbar\dot{\Phi}_{\iota}\left(t\right)-\Phi_{\iota}\left(t\right)\mathcal{L}_{S}\nonumber \\
 &  & \,+\frac{i}{\hbar}\int_{0}^{t}\mathrm{d}\tau\Phi_{\iota}\left(t-\tau\right)\kappa\left(\tau\right).\label{eq:kappa_I_EOM}
\end{eqnarray}
Once again, this is a closed form solution rather than an integral
equation. Its inputs are the same $\kappa\left(\tau\right)$ defined
in Eq.~\prettyref{eq:kappa_definition}, as well as the new quantity
\begin{eqnarray}
\Phi_{\iota}\left(t\right) & = & \mathrm{Tr}_{B}\left\{ I\mathcal{L}e^{-\frac{i}{\hbar}\mathcal{L}t}\rho_{B}\right\} .\label{eq:Phi_I_definition}
\end{eqnarray}

Eqs.~\prettyref{eq:kappa_I_EOM} and \prettyref{eq:iota_eom} amount
to a generalization of the Nakajima--Zwanzig--Mori formalism to non-system
operators. The structure of these equations is reminiscent of the
structure of the corresponding equations in the original theory, on
which the extension relies---and yet they are simpler in a certain
sense, as they are closed form solutions up to quadrature rather than
integro-differential or integral equations. In addition to $\sigma\left(t\right)$
and $\kappa\left(t\right)$, which can be obtained from the original
theory, the extended formalism relies on a new input, $\Phi_{\iota}\left(t\right)$,
which is defined in terms of regular (rather than projected) time
propagation and must be calculated explicitly. Once $\Phi_{\iota}\left(t\right)$
is available one can solve Eq.~\prettyref{eq:kappa_I_EOM} to obtain
$\kappa_{\iota}\left(t\right)$, and then solve Eq.~\prettyref{eq:iota_eom}
to obtain $\iota\left(t\right)$, a \textit{system-space} operator
which can be traced over to obtain the expectation value of the operator
$I$.

\section{Steady state\label{sec:Steady-State}}

If we wish to examine the $t\rightarrow\infty$ limit of $\sigma\left(t\right)$,
it is more convenient to define the Laplace transform 
\begin{equation}
\hat{\sigma}\left(z\right)=\int_{0}^{\infty}e^{-zt}\sigma\left(t\right)\,\mathrm{d}t
\end{equation}
When applied to Eq.~\prettyref{eq:sigma_eom}, this yields: 
\begin{eqnarray}
i\hbar\left[z\hat{\sigma}\left(z\right)-\sigma\left(0\right)\right] & = & \mathcal{L}_{S}\hat{\sigma}\left(z\right)+\hat{\vartheta}\left(z\right)-\frac{i}{\hbar}\hat{\kappa}\left(z\right)\hat{\sigma}\left(z\right),\\
 & \Downarrow\nonumber \\
\hat{\sigma}\left(z\right) & = & \frac{\sigma\left(0\right)+\frac{1}{i\hbar}\hat{\vartheta}\left(z\right)}{z-\frac{1}{i\hbar}\mathcal{L}_{S}+\frac{1}{\hbar^{2}}\hat{\kappa}\left(z\right)}.
\end{eqnarray}
Using the final value theorem $\sigma\left(\infty\right)=\lim_{z\rightarrow0}z\hat{\sigma}\left(z\right)$,
we can obtain an expression for $\hat{\sigma}$ at long times:
\begin{equation}
\sigma\left(t\rightarrow\infty\right)=\lim_{z\rightarrow0}\frac{i\hbar\sigma\left(0\right)+\hat{\vartheta}\left(z\right)}{\left(i\hbar+\frac{i}{\hbar}\frac{1}{z}\hat{\kappa}\left(z\right)-\frac{1}{z}\mathcal{L}_{S}\right)}.\label{eq:sigma_at_infinite_time}
\end{equation}
We can also obtain a stationary-state equation by considering a time
independent solution $\sigma\left(t\rightarrow\infty\right)$ to Eq.~\prettyref{eq:sigma_eom},
such that we can set the time derivative to zero and take $\sigma$
outside the integral before taking the Laplace transform. If we also
assume that the initial correlations are either zero to begin with
or die out at infinite time, this gives:
\begin{equation}
\left(\mathcal{L}_{S}-\frac{i}{\hbar}\hat{\kappa}\left(z\rightarrow i0\right)\right)\sigma\left(t\rightarrow\infty\right)=0.\label{eq:sigma_stationary_state}
\end{equation}
This last equation is of particular interest, because it allows us
to go from the memory kernel and system Liouvillian directly to the
steady state properties of the reduced density matrix, without passing
through the dynamics and without any reference to the initial state
or correlations of the system. This is very useful when we are interested
in general questions regarding the steady state, such as that of its
existence or uniqueness. When applying the cutoff approximation, $\hat{\kappa}\left(z\rightarrow i0\right)$
must be calculated to sufficient accuracy that the steady state density
matrix converges.

It is natural to attempt deriving a similar expression for the current
directly at steady state. One way of going about this task is to begin
with Eq.~\prettyref{eq:iota_eom} and take the Laplace transform:
\begin{eqnarray}
i\hbar\left[z\hat{\iota}\left(z\right)-\overset{{\scriptstyle =0}}{\overbrace{\iota\left(0\right)}}\right] & = & \mathcal{L}_{\iota}\hat{\sigma}\left(z\right)+\hat{\vartheta}_{\iota}\left(z\right)\nonumber \\
 &  & \,-\frac{i}{\hbar}\hat{\kappa}_{\iota}\left(z\right)\hat{\sigma}\left(z\right).
\end{eqnarray}
Extracting $z\hat{\iota}\left(z\right)$ and using the final value
theorem then gives
\begin{eqnarray}
\iota\left(t\rightarrow\infty\right) & = & \lim_{z\rightarrow0}\frac{1}{i\hbar}\left(\mathcal{L}_{\iota}-\frac{i}{\hbar}\hat{\kappa}_{\iota}\left(z\right)\right)\hat{\sigma}\left(z\right)\\
 & = & \lim_{z\rightarrow0}\frac{1}{i\hbar z}\left(\mathcal{L}_{\iota}-\frac{i}{\hbar}\hat{\kappa}_{\iota}\left(z\right)\right)\sigma\left(t\rightarrow\infty\right).
\end{eqnarray}
This suggests that in order for a steady state to exist, we must have
\begin{equation}
\lim_{z\rightarrow0}\mathcal{L}_{\iota}-\frac{i}{\hbar}\hat{\kappa}_{\iota}\left(z\right)\sim z.
\end{equation}
The constant of proportionality (itself a superoperator) determines
the value of the current at steady state. In the case of impurity
observables, it is sufficient to know the zero frequency component
of the memory kernel in order to obtain the steady-state value. Here,
however, one must also know something about the low-frequency properties
(or linear frequency response) of the current memory kernel, $\hat{\kappa}_{\iota}\left(z\right)$.

\section{Expressing the kernels in second-quantized form\label{sec:Expressing-kernels}}

Everything up to this point has been independent of the details of
any particular model, requiring only that a partitioning between the
impurity and bath part be made. In order to illustrate the process
of using the formalism presented above in a particular model, we will
continue by way of the simplest possible example: that of a noninteracting
junction (the formalism is not limited to this case\cite{cohen_memory_2011-1,cohen_numerically_2013,Wilner2013}).
This model, often called the resonant level model, is defined by the
Hamiltonian
\begin{eqnarray}
H & = & H_{S}+H_{L}+V,\label{eq:landauer_hamiltonian}\\
H_{S} & = & \varepsilon d^{\dagger}d,\label{eq:landauer_system_hamiltonian}\\
H_{L} & = & \sum_{q}\varepsilon_{q}a_{q}^{\dagger}a_{q},\label{eq:landauer_bath_hamiltonian}\\
V & = & \sum_{q}t_{q}da_{q}^{\dagger}+t_{q}^{*}a_{q}d^{\dagger}.\label{eq:landauer_coupling_hamiltonian}
\end{eqnarray}
A complete definition must include the $\varepsilon_{q}$and $t_{q}$,
and in this case all necessary information is contained in the lead
coupling function 
\begin{equation}
\Gamma\left(\omega\right)=2\pi\sum_{q}\left|t_{q}\right|^{2}\delta\left(\omega-\omega_{q}\right).\label{eq:Gamma_definition}
\end{equation}

The first step in the calculation is the evaluation of the system
Liovillian. It is convenient to work in the Hubbard representation
for operators in the impurity subspace: an operator $\hat{A}\in\mathcal{S}$
can be written as $\hat{A}=\sum_{ij}a_{ij}\left|i\right\rangle \left\langle j\right|$,
where the indices $i$ and $j$ can take on the values of states in
the impurity subspace---in this case $0$ and $1$ for unoccupied
and occupied, respectively. This superoperator simply performs a commutation
with the system Hamiltonian, and using Eq.~\prettyref{eq:landauer_system_hamiltonian}
to insert the explicit form of $H_{S}$ yields an expressions for
$\mathcal{L}_{S}$ in matrix (or tetradic) form:
\begin{eqnarray}
\left[\mathcal{L}_{s}\right]_{ij,kl} & = & \mathrm{Tr}_{S}\left\{ \left(\left|i\right\rangle \left\langle j\right|\right)^{\dagger}\mathcal{L}_{S}\left|k\right\rangle \left\langle l\right|\right\} \\
 & = & \sum_{m=0}^{1}\left\langle m\right|\left(\left|i\right\rangle \left\langle j\right|\right)^{\dagger}\left[\varepsilon\left|1\right\rangle \left\langle 1\right|,\left|k\right\rangle \left\langle l\right|\right]\left|m\right\rangle \\
 & = & \varepsilon\left[\delta_{jl0}\delta_{ik1}+\delta_{ijkl1}-\delta_{jl1}\delta_{ik}\right].\label{eq:landauer_system_liouvillian}
\end{eqnarray}
Here $\delta_{a_{1}a_{2}...a_{N}}$ is one if all indices take the
same value and zero otherwise. When no bath is present the model is
reduced to a two-level system, and Eq.~\prettyref{eq:sigma_eom}
gives the expected result:
\begin{eqnarray}
i\hbar\frac{\mathrm{d}\sigma_{ij}}{\mathrm{d}t} & = & \sum_{kl}\left[\mathcal{L}_{s}\right]_{ij,kl}\sigma_{kl}\\
 & = & \varepsilon\left(\delta_{i1}\delta_{j0}\sigma_{10}-\delta_{i0}\delta_{j1}\sigma_{01}.\right)
\end{eqnarray}
That is, off-diagonal density matrix elements oscillate with a frequency
$\frac{\varepsilon}{\hbar}$ while diagonal elements remain stationary.

Next, we need to evaluate the memory kernel. This requires the evaluation
of the superoperator 
\begin{eqnarray}
\Phi\left(t\right)A & = & \mathrm{Tr}_{B}\left\{ \mathcal{L}_{V}e^{-\frac{i}{\hbar}\mathcal{L}t}\rho_{B}A\right\} \\
 & = & \mathrm{Tr}_{B}\left\{ Ve^{-\frac{i}{\hbar}Ht}\rho_{B}Ae^{\frac{i}{\hbar}Ht}\right.\nonumber \\
 &  & \,\left.-e^{-\frac{i}{\hbar}Ht}\rho_{B}Ae^{\frac{i}{\hbar}Ht}V\right\} ,
\end{eqnarray}
which can also be represented in matrix form:
\begin{eqnarray}
\Phi_{ij,kl}\left(t\right) & = & \mathrm{Tr}_{S}\left\{ \left(\left|i\right\rangle \left\langle j\right|\right)^{\dagger}\phi\left(t\right)\left|k\right\rangle \left\langle l\right|\right\} \\
 & = & A-A^{\prime},
\end{eqnarray}
with
\begin{eqnarray}
A & \equiv & \mathrm{Tr}_{S}\left\{ \left|j\right\rangle \left\langle i\right|\mathrm{Tr}_{B}\left\{ Ve^{-\frac{i}{\hbar}Ht}\rho_{B}\left|k\right\rangle \left\langle l\right|e^{\frac{i}{\hbar}Ht}\right\} \right\} ,\\
A^{\prime} & \equiv & \mathrm{Tr}_{S}\left\{ \left|j\right\rangle \left\langle i\right|\mathrm{Tr}_{B}\left\{ e^{-\frac{i}{\hbar}Ht}\rho_{B}\left|k\right\rangle \left\langle l\right|e^{\frac{i}{\hbar}Ht}V\right\} \right\} .
\end{eqnarray}
Consider the term $A$. Let us perform the trace over the impurity
space and take the Hubbard operators to second quantized form: 
\begin{eqnarray}
A & = & \mathrm{Tr}_{B}\left\{ \rho_{B}\left\langle l\right|e^{\frac{i}{\hbar}Ht}\right.\nonumber \\
 &  & \,\left(\delta_{i1}\delta_{j1}d^{\dagger}d+\delta_{i0}\delta_{j0}dd^{\dagger}+\delta_{i1}\delta_{j0}d+\delta_{i0}\delta_{j1}d^{\dagger}\right)\nonumber \\
 &  & \,\left.Ve^{-\frac{i}{\hbar}Ht}\left|k\right\rangle \right\} \\
 & = & \mathrm{Tr}_{B}\Big\{\rho_{B}\left\langle l\right|\nonumber \\
 &  & \,\left(\delta_{i1}\delta_{j1}d^{\dagger}\left(t\right)d\left(t\right)+\delta_{i0}\delta_{j0}d\left(t\right)d^{\dagger}\left(t\right)\right.\nonumber \\
 &  & \,\left.+\delta_{i1}\delta_{j0}d\left(t\right)+\delta_{i0}\delta_{j1}d^{\dagger}\left(t\right)\right)\nonumber \\
 &  & \, V\left(t\right)\left|k\right\rangle \Big\}.
\end{eqnarray}
In the final step, the operators were given their full time dependence
in the Heisenberg picture. Using $V\left(t\right)=\sum_{q}t_{q}d\left(t\right)a_{q}^{\dagger}\left(t\right)+t_{q}^{*}a_{q}\left(t\right)d^{\dagger}\left(t\right)$
and the fact that all pairs of dot and lead operators maintain normal
commutation relations when taken at the same times, one can now show
that 
\begin{eqnarray}
A & = & \sum_{q}t_{q}\mathrm{Tr}_{B}\left\{ \rho_{B}\left\langle l\right|\delta_{i0}\delta_{j0}d\left(t\right)a_{q}^{\dagger}\left(t\right)\left|k\right\rangle \right\} \\
 &  & \,+\sum_{q}t_{q}\mathrm{Tr}_{B}\left\{ \rho_{B}\left\langle l\right|\delta_{i0}\delta_{j1}d^{\dagger}\left(t\right)d\left(t\right)a_{q}^{\dagger}\left(t\right)\left|k\right\rangle \right\} \nonumber \\
 &  & \,-\sum_{q}t_{q}^{*}\mathrm{Tr}_{B}\left\{ \rho_{B}\left\langle l\right|\delta_{i1}\delta_{j1}d^{\dagger}\left(t\right)a_{q}\left(t\right)\left|k\right\rangle \right\} \nonumber \\
 &  & -\sum_{q}t_{q}^{*}\mathrm{Tr}_{B}\left\{ \rho_{B}\left\langle l\right|\delta_{i1}\delta_{j0}d\left(t\right)d^{\dagger}\left(t\right)a_{q}\left(t\right)\left|k\right\rangle \right\} .\nonumber 
\end{eqnarray}
Similarly,
\begin{eqnarray}
A^{\prime} & = & \sum_{q}t_{q}\mathrm{Tr}_{B}\left\{ \rho_{B}\left\langle l\right|\delta_{i1}\delta_{j1}d\left(t\right)a_{q}^{\dagger}\left(t\right)\left|k\right\rangle \right\} \\
 &  & \,-\sum_{q}t_{q}\mathrm{Tr}_{B}\left\{ \rho_{B}\left\langle l\right|\delta_{i0}\delta_{j1}d\left(t\right)d^{\dagger}\left(t\right)a_{q}^{\dagger}\left(t\right)\left|k\right\rangle \right\} \nonumber \\
 &  & \,-\sum_{q}t_{q}^{*}\mathrm{Tr}_{B}\left\{ \rho_{B}\left\langle l\right|\delta_{i0}\delta_{j0}d^{\dagger}\left(t\right)a_{q}\left(t\right)\left|k\right\rangle \right\} \nonumber \\
 &  & \,+\sum_{q}t_{q}^{*}\mathrm{Tr}_{B}\left\{ \rho_{B}\left\langle l\right|\delta_{i1}\delta_{j0}d^{\dagger}\left(t\right)d\left(t\right)a_{q}\left(t\right)\left|k\right\rangle \right\} .\nonumber 
\end{eqnarray}
Putting the expressions for $A$ and $A^{\prime}$ into their defining
equation then yields
\begin{eqnarray}
\Phi_{ij,kl}\left(t\right) & = & -2i\left(\delta_{i1}\delta_{j1}-\delta_{i0}\delta_{j0}\right)\Im\left\{ \varphi_{kl}\right\} \nonumber \\
 &  & +\delta_{i0}\delta_{j1}\psi_{kl}-\delta_{i1}\delta_{j0}\psi_{lk}^{*},\label{eq:phi_matrix-1}
\end{eqnarray}
with
\begin{eqnarray}
\varphi_{kl} & = & \mathrm{Tr}_{B}\left\{ \sum_{q}t_{q}\rho_{B}\left\langle l\right|d\left(t\right)a_{q}^{\dagger}\left(t\right)\left|k\right\rangle \right\} ,\label{eq:varphi_kl}\\
\psi_{kl} & = & \mathrm{Tr}_{B}\left\{ \sum_{q}t_{q}\rho_{B}\left\langle l\right|a_{q}^{\dagger}\left(t\right)\left|k\right\rangle \right\} .\label{eq:varpsi_kl}
\end{eqnarray}
The $\varphi$ elements have a rather simple physical interpretation:
they are directly proportional to the time derivative of the total
population on the dot. The $\psi$ elements are harder to interpret
in such a manner.

The equations therefore collapse to a simple form, phrased in terms
of the system-space matrix elements $\varphi_{kl}$ and $\psi_{kl}$
of normal second quantization operators propagated under the influence
of the full Hamiltonian. Some further simplification can be made by
considering Eq.~\prettyref{eq:varphi_kl} as a function of time when
we go to the interaction picture under $H_{0}=H_{S}+H_{B}$:
\begin{eqnarray}
\varphi_{kl} & = & \mathrm{Tr}_{B}\left\{ \sum_{q}t_{q}\rho_{B}\left\langle l\right|e^{iHt}da_{q}^{\dagger}e^{-iHt}\left|k\right\rangle \right\} \\
 & = & \sum_{q}t_{q}\mathrm{Tr}_{B}\nonumber \\
 &  & \,\left\{ \rho_{B}\left\langle l\right|U^{\dagger}\left(t\right)d_{H_{0}}\left(t\right)a_{H_{0},q}^{\dagger}\left(t\right)U\left(t\right)\left|k\right\rangle \right\} .
\end{eqnarray}
It is easy to verify that the time dependence of $d$ and $a_{q}^{\dagger}$
in the interaction-picture is described by a simple oscillation, and
that the $U$ and $U^{\dagger}$ are sums over products of terms containing
either $a_{q}^{\dagger}d$ or $d^{\dagger}a_{q}$ in the interaction
picture. The trace over the bath may then be performed at time zero,
throwing out all terms which do not have the same number of $a_{q}$
and $a_{q}^{\dagger}$ operators. Yet, from the argument we have just
stated, such terms will also have the same number of $d$ and $d^{\dagger}$
operators, and $\varphi_{kl}$ must be zero unless $k=l$. For similar
considerations, $\psi_{kl}$ is zero unless $k\neq l$ and we have
\begin{eqnarray}
\varphi_{kl} & = & \mathrm{Tr}_{B}\left\{ \sum_{q}t_{p}\rho_{B}\left\langle p\right|d\left(t\right)a_{q}^{\dagger}\left(t\right)\left|k\right\rangle \right\} \delta_{kl},\\
\psi_{kl} & = & \mathrm{Tr}_{B}\left\{ \sum_{q}t_{q}\rho_{B}\left\langle l\right|a_{q}^{\dagger}\left(t\right)\left|k\right\rangle \right\} \left(1-\delta_{kl}\right).
\end{eqnarray}
Considering the role of $\varphi$ and $\psi$ in Eq.~\prettyref{eq:phi_matrix-1},
one can see that $\Phi$ contains terms which couple the populations
and terms which couple the coherences; it contains no terms which
couple the populations to the coherences. Examining Eq.~\prettyref{eq:kappa_EOM},
one realizes that terms with no inhomogeneous contribution must identically
vanish, such that
\begin{equation}
\kappa_{ij,kl}=\begin{cases}
\kappa_{ii,kk}: & i=j,\, k=l,\\
\kappa_{ij,kl}: & i\neq j,\, k\neq l,\\
0: & \mathrm{otherwise}.
\end{cases}
\end{equation}
Similarly, Eq.~\prettyref{eq:sigma_eom}, along with the Liouvillian
\prettyref{eq:landauer_system_liouvillian}, immediately leads us
to the conclusion that the diagonal elements of $\sigma$ form one
coupled block within the formalism, while the off-diagonal elements
of $\sigma$ form a second block: in other words, within the resonant
level model, the reduced dynamics of the diagonal elements (the populations)
are decoupled from those of the off-diagonal elements (the coherences). 

Among other things, this implies that if we are interested only in
the populations we do not need to calculate the $\psi_{kl}$, and
vice-versa for the coherences. Since the populations are also unaffected
by the Liouvillian, and assuming factorized initial conditions, the
equation of motion turns from a superoperator equation into a matrix
equation for the population vector $\sigma_{ii}$:
\begin{equation}
i\hbar\frac{\mathrm{d}}{\mathrm{d}t}\sigma_{ii}\left(t\right)=-\frac{i}{\hbar}\int_{0}^{t}\mathrm{d}\tau\,\sum_{j}\kappa_{ii,jj}\left(\tau\right)\sigma_{jj}\left(t-\tau\right).
\end{equation}
Interestingly, this analytic block structure conclusion holds in the
generalized Holstein model~\cite{Wilner2013} as well (though this
will not be shown here), and continues to hold for the Anderson model~\cite{cohen_memory_2011-1}
even in the presence of a magnetic field.\cite{cohen_numerically_2013}

We continue to examine the memory formalism for the non-system current
operator in the noninteracting case. We will define a simplified left
current operator as 
\begin{equation}
\tilde{I}\equiv\sum_{q\in L}t_{q}da_{q}^{\dagger},
\end{equation}
such that $\tilde{\iota}\left(t\right)=\mathrm{Tr}_{B}\left\{ \tilde{I}\rho\left(t\right)\right\} $
and the physical current is 
\begin{equation}
\left\langle I\left(t\right)\right\rangle =-2\frac{e}{\hbar}\Im\left\langle \tilde{I}\left(t\right)\right\rangle =-2\frac{e}{\hbar}\Im\mathrm{Tr}_{S}\left\{ \tilde{\iota}\left(t\right)\right\} .\label{eq:physical_current_from_iota_tilde}
\end{equation}
First, consider the current Liouvillian operator:
\begin{eqnarray}
\mathrm{Tr}_{B}\left\{ \tilde{I}\mathcal{L}\rho_{B}\right\}  & = & \mathrm{Tr}_{B}\left\{ \tilde{I}\mathcal{L}\rho_{B}\right\} \\
 & = & \mathrm{Tr}_{B}\left\{ \sum_{q\in L}t_{q}da_{q}^{\dagger}\mathcal{L}_{V}\rho_{B}\right\} .
\end{eqnarray}
The term containing\emph{ $\mathcal{L}_{B}$} can be dropped because
$\mathcal{L}_{B}\rho_{B}A=\left(\mathcal{L}_{B}\rho_{B}\right)A=0$
for any system variable $A$, while the term containing\emph{ $\mathcal{L}_{S}$}
is zero because it must contain a trace over an odd number of lead
creation or destruction operators. It is then simple to show by the
same procedure from which the system Liouvillian was derived that
\begin{eqnarray}
\left[\mathcal{L}_{\iota}\right]_{ij,ml} & = & \mathrm{Tr}_{S}\left\{ \left(\left|i\right\rangle \left\langle j\right|\right)^{\dagger}\mathrm{Tr}_{B}\left\{ \tilde{I}\mathcal{L}\rho_{B}\right\} \left|m\right\rangle \left\langle l\right|\right\} \\
 & = & \sum_{q\in L}\left|t_{q}\right|^{2}\mathrm{Tr}\left\{ \left(\left|i\right\rangle \left\langle j\right|\right)^{\dagger}dd^{\dagger}a_{q}^{\dagger}a_{q}\rho_{B}\left|m\right\rangle \left\langle l\right|\right\} \\
 &  & \,-\sum_{q\in L}\left|t_{q}\right|^{2}\mathrm{Tr}\left\{ \left(\left|i\right\rangle \left\langle j\right|\right)^{\dagger}da_{q}^{\dagger}\rho_{B}\left|m\right\rangle \left\langle l\right|a_{q}d^{\dagger}\right\} \nonumber \\
 & = & \sum_{q\in L}\left|t_{q}\right|^{2}\mathrm{Tr}_{B}\left\{ \left\langle i\right|dd^{\dagger}f_{q}\rho_{B}\left|m\right\rangle \left.\left\langle l\right|j\right\rangle \right\} \\
 &  & \,-\sum_{q\in L}\left|t_{q}\right|^{2}\mathrm{Tr}_{B}\left\{ \left\langle i\right|da_{q}^{\dagger}\rho_{B}\left|m\right\rangle \left\langle l\right|a_{q}d^{\dagger}\left|j\right\rangle \right\} ,\nonumber 
\end{eqnarray}
where $f_{q}=\frac{1}{1+e^{\beta(\varepsilon_{q}-\mu_{L})}}$ is the
Fermi-Dirac distribution. In the above equation, we have used the
factorized initial conditions by assuming an equilibrium Fermi distribution
at $t=0$ in the baths (it is worth noting that this distribution
is allowed to evolve freely under the influence of the full Hamiltonian
in the reduced dynamics formalism, yet the full details of bath dynamics
are no longer accessible from the information stored in $\sigma\left(t\right)$).
With this, it is straightforward to show that
\begin{eqnarray}
\left[\mathcal{L}_{\iota}\right]_{ij,ml} & = & \delta_{i0}\delta_{m0}\delta_{lj}\Delta^{<}\left(0\right)-\delta_{i0}\delta_{j0}\delta_{m1}\delta_{l1}\Delta^{>}\left(0\right),
\end{eqnarray}
where the hybridization functions
\begin{eqnarray}
\Delta^{<}\left(t\right) & = & \int\mathrm{d}\omega\, e^{i\omega t}\sum_{q\in L}\left|t_{q}\right|^{2}f_{q}\delta\left(\omega-\omega_{q}\right),\\
\Delta^{>}\left(t\right) & = & \int\mathrm{d}\omega\, e^{i\omega t}\sum_{q\in L}\left|t_{q}\right|^{2}\left(1-f_{q}\right)\delta\left(\omega-\omega_{q}\right),
\end{eqnarray}
can easily be evaluated in terms of the frequency-space coupling density
given in Eq.~\prettyref{eq:Gamma_definition}.

The final object we need to evaluate is $\Phi_{\iota}$. Since no
new conceptual issues arise here as compared with the calculation
performed for $\Phi$, we will simply write down the final answer.
With the definitions
\begin{eqnarray}
\varphi_{\iota,1}^{<} & \equiv & \sum_{q\in L,q^{\prime}}t_{q}t_{q^{\prime}}^{*}\left\langle a_{q}^{\dagger}\left(t\right)a_{q^{\prime}}\left(t\right)\right\rangle \\
 &  & \,-\sum_{q\in L}\left(\left|t_{q}\right|^{2}\left\langle d^{\dagger}\left(t\right)d\left(t\right)\right\rangle +t_{q}\varepsilon_{q}\left\langle d\left(t\right)a_{q}^{\dagger}\left(t\right)\right\rangle \right),\nonumber \\
\varphi_{\iota,2} & \equiv & \sum_{q\in L}t_{q}\left\langle d\left(t\right)a_{q}^{\dagger}\left(t\right)\right\rangle ,\\
\psi_{\iota,1} & \equiv & \sum_{q\in L,q^{\prime}}t_{q}t_{q^{\prime}}^{*}\left\langle d^{\dagger}\left(t\right)a_{q}^{\dagger}\left(t\right)a_{q^{\prime}}\left(t\right)\right\rangle ,\\
\psi_{\iota,2} & \equiv & \sum_{q\in L,q^{\prime}}t_{q}t_{q^{\prime}}\left\langle da_{q^{\prime}}^{\dagger}\left(t\right)a_{q}^{\dagger}\left(t\right)\right\rangle ,
\end{eqnarray}
$\Phi_{\iota}\left(t\right)$ takes the simple form:
\begin{eqnarray}
\left[\Phi_{\iota}\right]_{ij,kl}\left(t\right) & = & \left\langle l\right|\left\{ \delta_{i0}\delta_{j0}\left(\varepsilon\varphi_{\iota,2}+\varphi_{\iota,1}^{<}\right)\right.\nonumber \\
 &  & \,\left.+\delta_{i0}\delta_{j1}\left(\psi_{\iota,1}-\psi_{\iota,2}\right)\right\} \left|k\right\rangle .
\end{eqnarray}
The inherent asymmetry of the expression above is due to the asymmetric
definition of $\tilde{I}$ (a symmetric definition would have generated
nonzero matrix elements at $i=j=1$ and at $i=1,\, j=0$, and thus
our choice was motivated by computational economy). As before, it
is easy to show that the block structure is such that the current
can be determined without reference to the off-diagonal elements of
either $\sigma(t)$ or $\iota(t)$. Furthermore, it is straightforward
to show that for the resonant level model (and for the Holstein model)$\left[\Phi_{\iota}\right]_{00,00}\left(t\right)=\frac{d\tilde{I}}{dt}(0)$
for an initially empty dot and $\left[\Phi_{\iota}\right]_{00,11}\left(t\right)=\frac{d\tilde{I}}{dt}(1)$
for an initially occupied dot. These are useful relations as they
provide an alternative way of computing $\kappa_{\iota}\left(t\right)$
directly from the left current (at short times), without the need
to evaluate $\varphi_{\iota,2}(t)$ or $\varphi_{\iota,1}^{<}(t)$.

\section{Results\label{sec:Results}}

\begin{figure}
\includegraphics[width=8.6cm]{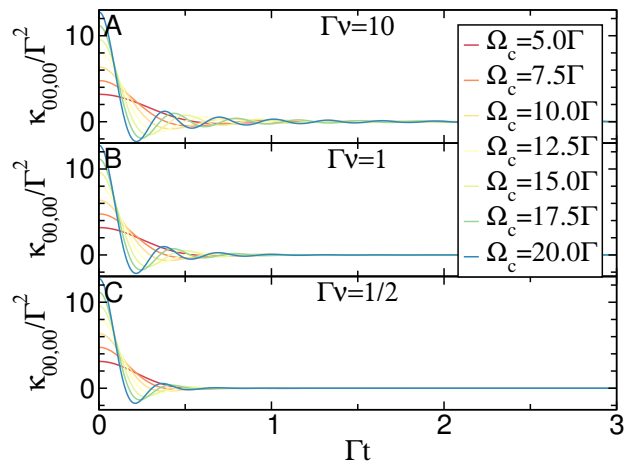}\caption{An element of the memory kernel $\kappa$ of the fully symmetric,
resonant level model at a range of band parameters. Due to the symmetry,
the results shown here are independent of both temperature and voltage.
Panel A through C show the effect of softening the band edge by varying
$\nu$, while within each separate panel the effect of varying the
bandwidth (which is approximately twice the cutoff frequency $\Omega_{C}$)
is illustrated.\label{fig:kappa_BW}}
\end{figure}

The physics of the resonant level model are generally well known,
yet the literature has seen little exploration of the properties of
the memory kernel in this model, and of course none of the current
memory kernel which has been introduced here. We therefore present
some results below which we expect to be of interest to the field,
as they provide insight into those aspects of the problem which do
not rely on interaction. In order to restrict the parameter space
explored, we will discuss the symmetric case in which $\varepsilon=0$
with a bias voltage applied symmetrically such that $V=2\mu_{L}=-2\mu_{R}$
(from here on we set $\hbar=e=1$). The lead coupling densities are
taken to be $\Gamma_{L,R}\left(\omega\right)=\frac{1}{\left(1+e^{\nu\left(\omega-\Omega_{C}\right)}\right)\left(1+e^{\nu\left(-\omega-\Omega_{C}\right)}\right)}$.
We will limit our attention only to the diagonal elements of the reduced
density matrix and the corresponding element of the memory kernel;
these elements are completely decoupled from the off-diagonal coherences,
as discussed above, and therefore no approximation ensues from this. 

\begin{figure}
\includegraphics[width=8.6cm]{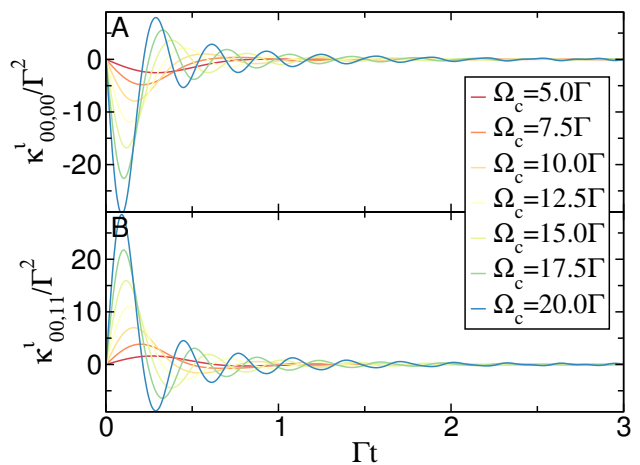}\caption{The two nonzero elements of $\kappa_{\iota}$ for the left current
are shown in panels A and B, with $\Gamma\nu=10$, $\Gamma\beta=1$
and $V=4\Gamma$. In each panel, the time dependence of the $\kappa_{\iota}$
element is shown at a range of bandwidths.\label{fig:kappaI_BW_nu10.0}}
\end{figure}

\begin{figure}
\includegraphics[width=8.6cm]{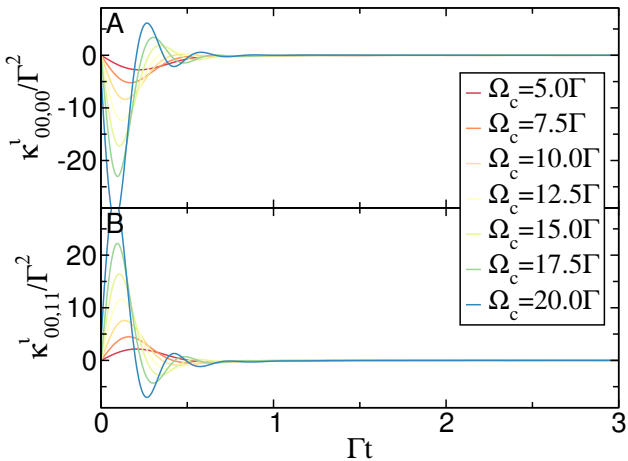}\caption{The two nonzero elements of $\kappa_{\iota}$ for the left current
are shown in panels A and B, with $\Gamma\nu=0.5$, $\Gamma\beta=1$
and $V=4\Gamma$. In each panel, the time dependence of the $\kappa_{\iota}$
element is shown at a range of bandwidths.\label{fig:kappaI_BW_nu0.5}}
\end{figure}

All the results presented in this section are exact and have been
calculated by a direct solution of the full equation of motion of
the complete density matrix, a technique which relies on the quadratic
form of the Hamiltonian and is therefore applicable only to the noninteracting
case. In general, making similar progress for interacting systems
requires a numerical solver of one type or another.\cite{muhlbacher_real-time_2008-1,weiss_iterative_2008,werner_diagrammatic_2009,wang_numerically_2009-1,segal_numerically_2010-1}

We begin with a discussion of the behavior of the memory kernel. The
symmetrical parameters we have chosen are of particular interest because
in the absence of interaction both $\sigma\left(t\right)$ and $\kappa\left(t\right)$
are completely independent of both the temperature and voltage. In
addition, all nonzero matrix elements of $\kappa$ are all identical
to within a sign ($\kappa_{00,00}=\kappa_{11,11}=-\kappa_{11,00}=-\kappa_{00,11}$).
In Fig.~\ref{fig:kappa_BW} we therefore explore the dependence of
one arbitrarily chosen element of $\kappa$ on a range of band parameters:
cutoff energies $\Omega_{C}$ (the bandwidth is $\sim2\Omega_{C}$)
and band cutoff widths $\frac{1}{\nu}$. In each panel we go from
a small bandwidth (red) to a large one (blue) at a set cutoff width,
with the sharpest cutoff shown in panel A, an intermediate value in
B and the smoothest in C.

The effect of the the two parameters describing our chosen band shape
on the memory kernel can be understood quite well by considering the
trends shown in the plot: as $\nu$ decreases, reflections are softened
by the gradual slope at the band edge and the memory kernel decays
more quickly. On the other hand, increasing the bandwidth induces
oscillations at a frequency $\omega\approx\Omega_{C}$, but also increases
the proportional weight of the short-time part of the memory kernel.
Eventually, if we were to approach the wide band limit, the memory
would approach the form of a delta function and a Markovian description
of the dynamics would become exact.

The current memory kernel $\kappa_{\iota}$ is not as highly symmetric
as $\kappa$, and depends to some extent on all the parameters of
the problem. There are two distinct (though similar) elements in $\kappa_{\iota}$
at nonzero voltage, and these are plotted for the left current with
$\Gamma\nu=10$ in panels A and B of Fig~\ref{fig:kappaI_BW_nu10.0}.
The figure illustrates the dependence of $\kappa_{\iota}$ on the
bandwidth, which exhibits the same properties observed in $\kappa$.
This is also true of its $\nu$ dependence: to exemplify this, Fig~\ref{fig:kappaI_BW_nu0.5}
displays the same data for $\Gamma\nu=0.5$, where $\kappa_{\iota}$
decays more quickly and smoothly. Interestingly, the timescale over
which $\kappa_{\iota}$ decays to zero does not appear to differ markedly
from the corresponding timescale for $\kappa$ at similar parameters;
this suggests that the cutoff approximation remains as useful for
the current as it is for impurity observables.

\begin{figure}
\includegraphics[width=8.6cm]{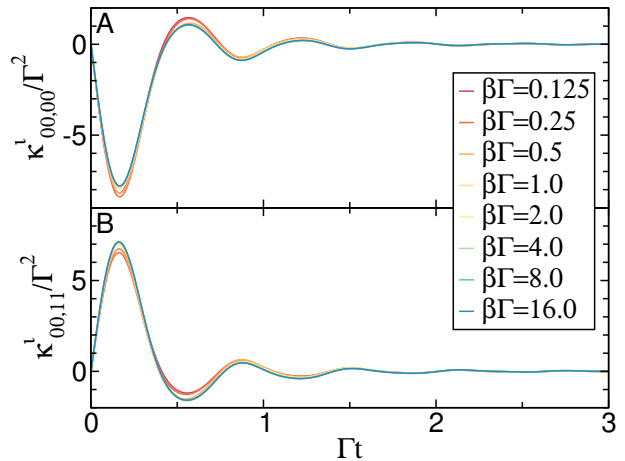}\caption{The two nonzero elements of $\kappa_{\iota}$ are shown in panels
A and B, with $\Gamma\nu=10$, $\Omega_{c}=10\Gamma$ and $V=4\Gamma$.
In each panel, the time dependence of the $\kappa_{\iota}$ element
is shown at a range of inverse temperatures $\beta$.\label{fig:kappaI_beta}}
\end{figure}

The effect of temperature on $\kappa_{\iota}$ depends greatly on
the choice of other parameters. At the parameters we have chosen for
Fig.~\ref{fig:kappaI_beta} the asymmetry between the two $\kappa_{\iota}$
elements is increased somewhat when the temperature is lowered, corresponding
to an increase in the current (not shown). One would expect a rather
different effect when, for instance, things are set up in such a way
that thermal enhancement of the current occurs. Unlike $\kappa$,
$\kappa_{\iota}$ depends on temperature even in the fully symmetric
and noninteracting case is interesting, and expresses the fact that
this quantity is connected to bath observables as well as to those
in the impurity subspace.

\begin{figure}
\includegraphics[width=8.6cm]{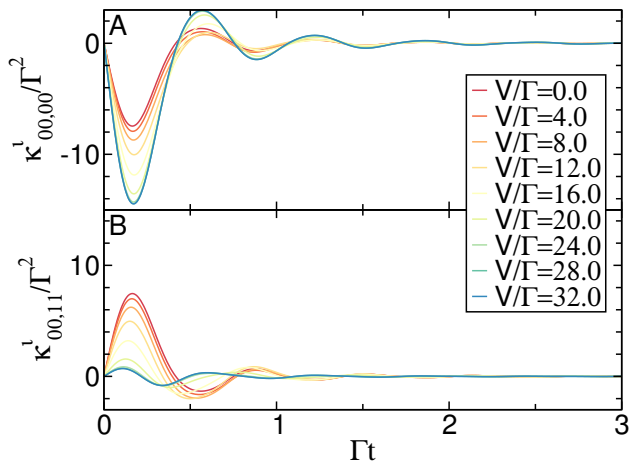}\caption{The two nonzero elements of $\kappa_{\iota}$ are shown in panels
A and B, with $\Gamma\nu=10$, $\Omega_{c}=10\Gamma$ and $\Gamma\beta=1$.
In each panel, the time dependence of the $\kappa_{\iota}$ element
is shown at a range of bias voltages $V$.\label{fig:kappaI_V}}
\end{figure}

Fig.~\ref{fig:kappaI_V} shows how voltage affects the current memory
kernel: at zero voltage the two elements of $\kappa_{\iota}$ are
identical up to a sign, and the application of a voltage increases
the diagonal element while suppressing the off-diagonal terms. Additionally,
an increase in the oscillation frequency is observed for $\kappa_{00,11}$,
but no such clear trend exists for the oscillation frequency in $\kappa_{00,00}$.
As $V$ passes the bandwidth ($\sim20\Gamma$ here), the left lead
becomes entirely occupied and the right entirely empty, and further
increasing the voltage ceases to have any effect on the current memory
kernel, just as occurs in the case of the current itself (not shown).

\begin{figure}
\includegraphics[width=8.6cm]{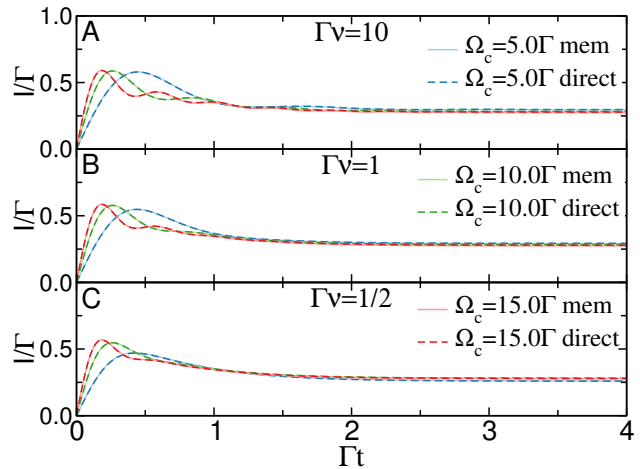}\caption{The left current is shown as a function of the physical time $t$
at a variety of band parameters. Dashed lines in dark colors correspond
to exact results calculated directly. Each such line is paired with
a solid line in a brighter color at the same parameters showing converged
results obtained from tracing over the $\iota$ operator obtained
from solving the generalized NZME Eq.~\prettyref{eq:iota_eom}. In
all cases we have set $\Gamma\beta=1$ and $V=4\Gamma$.\label{fig:current_examples}}
\end{figure}

To show that the generalized NZME formalism introduced in this work
indeed reproduces the correct results for the current as a function
of time, Fig.~\ref{fig:current_examples} presents a comparison between
the current obtained directly (lighter solid lines) and by way of
Eq.~\prettyref{eq:iota_eom} (darker dashed lines). Pairs of lines
describing the two different ways of obtaining currents at identical
parameters overlap to within numerical errors, expressing the equivalence
between the two methods when convergence in the cutoff time has been
attained and the correctness of the approach at the $t_{c}\rightarrow\infty$
limit.

\begin{figure}
\includegraphics[width=8.6cm]{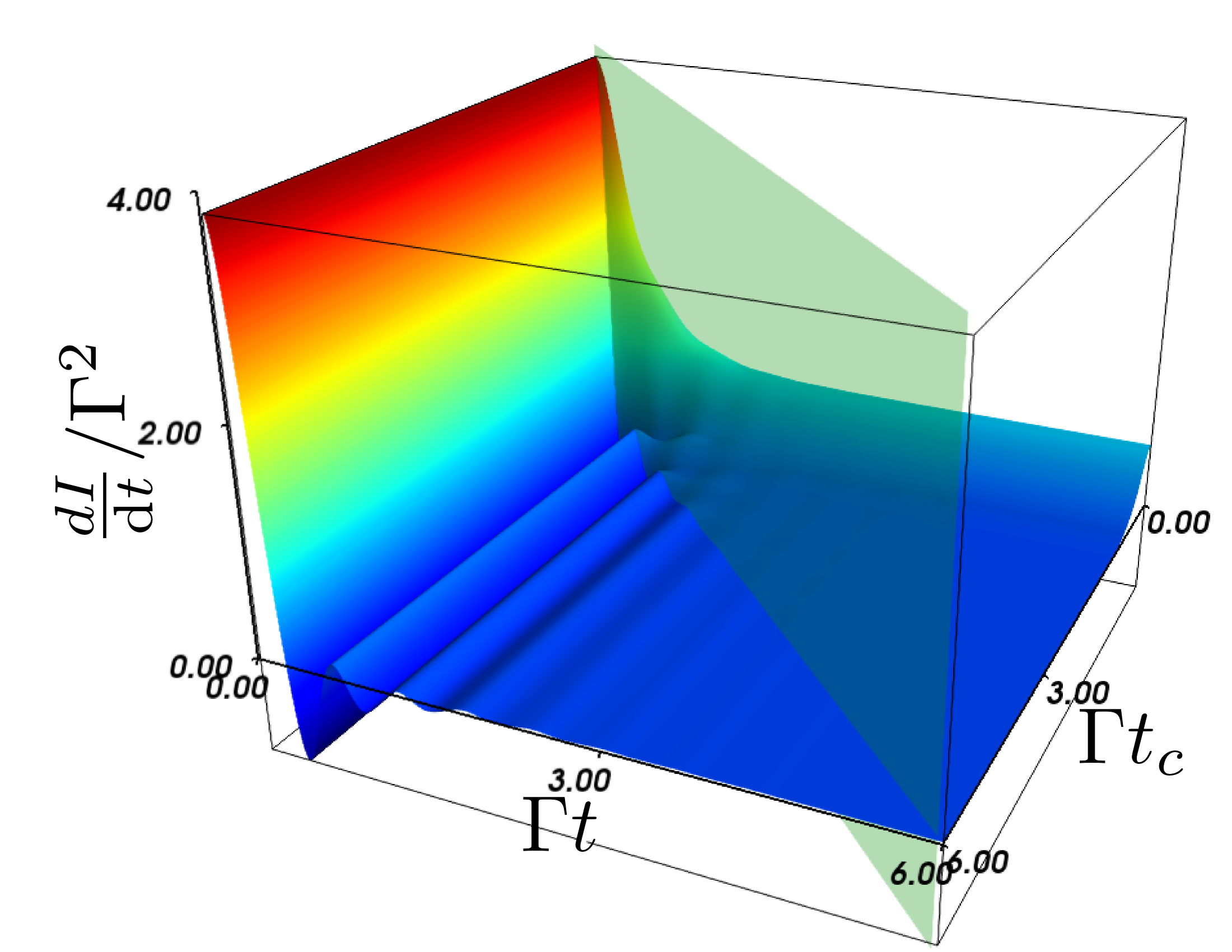}\caption{The time derivative of the left current in the cutoff approximation
is shown as a function of both physical time $t$ and the cutoff time
$t_{c}$. The transparent plane marks $t=t_{c}$, and for $t<t_{c}$
(to the left of the plane) the results are exact. To converge the
results for the current within some numerical accuracy, one must increase
$t_{c}$ until $\frac{\mathrm{dI\left(t\right)}}{\mathrm{d}t}$ ceases
to vary within that accuracy. Parameters are $\Gamma\nu=10$, $\Omega_{c}=10\Gamma$,
$\Gamma\beta=1$ and $V=4\Gamma$. \label{fig:diota_convergence}}
\end{figure}

\begin{figure}
\includegraphics[width=8.6cm]{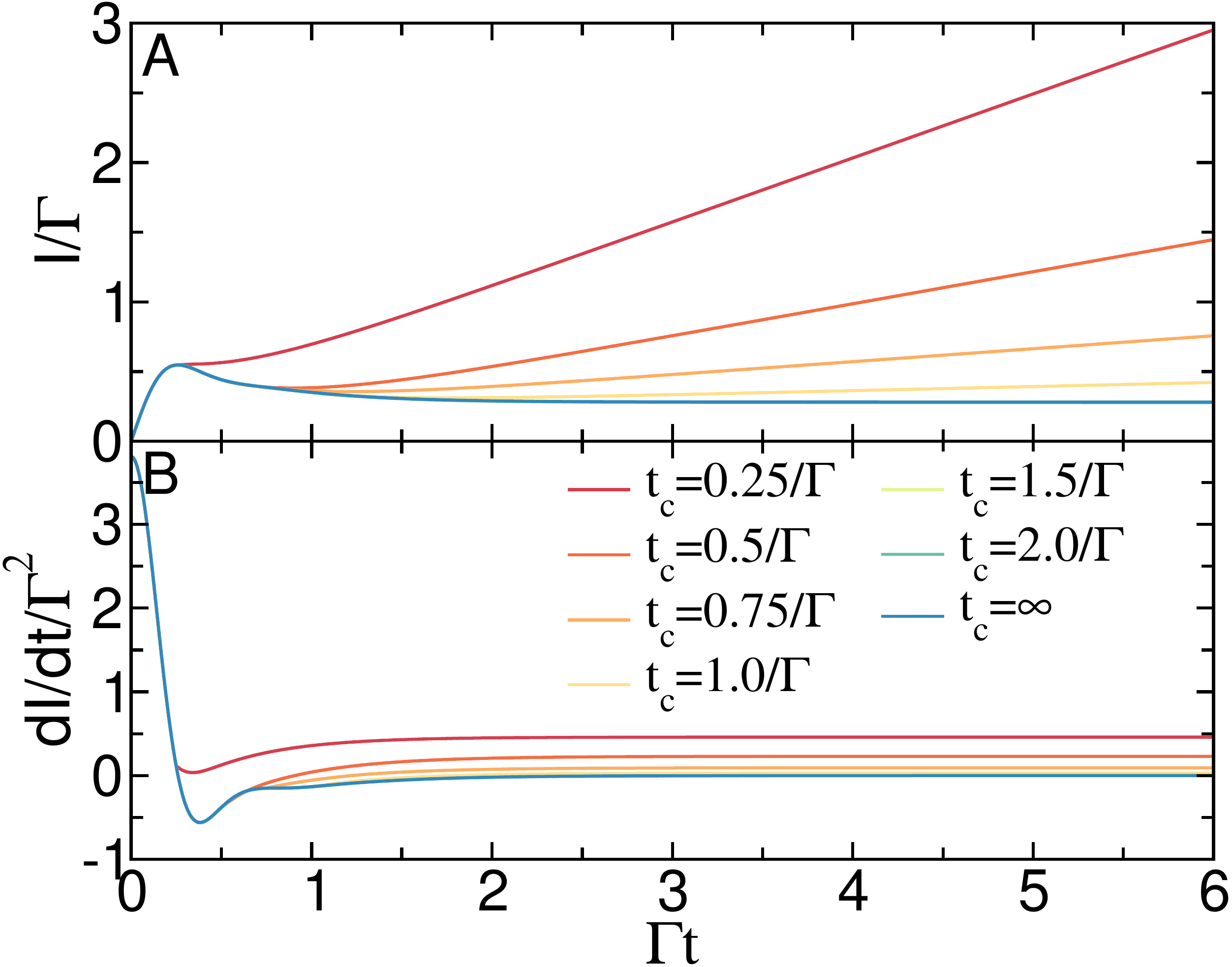}\caption{The left current (top) and its time derivative (bottom) are shown
as a function of the physical time $t$ at $\Gamma\nu=0.5$, $\Omega_{c}=10\Gamma$,
$\Gamma\beta=1$ and $V=4\Gamma$, for a range of cutoff times $t_{c}$.
The final line, labeled $t_{c}=\infty$, shows the exact result for
comparison.\label{fig:current_cutofftimes}}
\end{figure}

Finally, while the NZME memory kernel technique and the cutoff approximation
have been shown to be efficient for $\sigma$ in a variety of interacting
and noninteracting cases,\cite{cohen_memory_2011-1,cohen_numerically_2013,Wilner2013}
meaning that results at long times $t\gg t_{c}$ converge at a finite
$t_{c}$, no such calculations have previously been carried out for
the generalized technique. This entails a convergence analysis of
the type exemplified graphically in Fig.~\ref{fig:diota_convergence}.
In essence, the cutoff time $t_{c}$ must be increased until the desired
accuracy is reached. In the example shown in Fig.~\ref{fig:diota_convergence},
convergence is achieved quickly and even short-time oscillations beyond
the range of $t_{c}$ are predicted with some accuracy (as can be
seen from the extension of the oscillatory ridges beyond the boundary
of the transparent $t=t_{c}$ plane). As an alternative (if partial)
representation of this idea, in Fig.~\ref{fig:current_cutofftimes}
several plane cuts through this function are shown, but this time
at $\Gamma\nu=0.5$; both the current and its time derivative are
displayed. The rapid convergence visible in either representation
illustrates that the idea of the reduced dynamics technique and the
cutoff approximation remains useful in practice even for the current,
despite it being a non-system operator not accessible within the confines
of the standard NZME formalism. 

\section{Summary and conclusions\label{sec:Summary-and-conclusions}}

We have reviewed the process of implementing reduced dynamics techniques
by way of the NZME and the memory cutoff approximation, which have
recently been introduced with great success into several numerically
exact nonequilibrium impurity solvers. The procedure of deriving a
memory kernel scheme for a general impurity model and obtaining calculable
expressions in terms of standard second-quantization operators was
outlined, and the example of the noninteracting resonant level model
was worked out in full detail. For this noninteracting case, some
illustrative examples of the physical properties of the memory kernel
were discussed.

An important limitation of the reduced dynamics techniques so far
has been the lack of access to none-impurity observables, such as
the electronic current in the resonant level model, the Anderson impurity
model, and the Holstein model. A generalization of the NZME formalism
which allows access to general operators was therefore introduced
here, and the implementation of this formalism was carried through
for the example of the current in the non-interacting limit. This
led to the definition and evaluation of a current memory kernel $\kappa_{\iota}$,
which was subsequently explored for its dependence on time, bandwidth,
voltage and temperature. The validity of the cutoff approximation
for the current memory kernel was then verified and discussed.

Looking forward, we expect the ideas expounded upon here to have several
major implications: first, we hope to see them become a standard part
of the toolbox of high quality time domain numerical simulations of
impurity models, and extended to a variety of models and methods;
in this context the memory technique should be viewed not as competing
with existing direct solvers, but as a supplemental tool which allows
efficient extension of any general short-time solver to long timescales,
in situations where the memory timescale is short. Second, since access
to current enables access to Green's functions, the benefits offered
by memory techniques are expected to be applicable to interacting
lattice simulations as well, by way of mapping schemes such as dynamical
mean field theory and its various extensions. Finally, we believe
the memory kernel framework is a fertile ground for defining new approximation
schemes more general than the cutoff approximation, and in this context
it will be particularly interesting to understand the long-time behavior
of the memory kernel in interacting cases and its behavior in larger
impurity models. 
\begin{acknowledgments}
\textcolor{black}{The authors would like to thanks A. Nitzan and M.R.
Wegewijs for insightful comments and helpful conversations. GC is
grateful to Yad Hanadiv--Rothschild Foundation for the award of a
Rothschild Postdoctoral Fellowship. EYW is grateful to The Center
for Nanoscience and Nanotechnology at Tel Aviv University for a doctoral
fellowship. This work was supported by the US--Israel Binational Science
Foundation.}
\end{acknowledgments}
\bibliographystyle{apsrev}
\bibliography{Library}

\end{document}